# Comment on "Fluctuation-dissipation considerations and damping models for ferromagnetic thin films"

Neil Smith
San Jose Research Center
Hitachi Global Storage Technologies
San Jose, CA 95120

In a recent article (Phys. Rev. B **71**, 224402 (2005)), Safonov and Bertram claim "inconsistent" results can occur when applying the "Callen-Welton fluctuation dissipation theorem" to magnetic systems with dissipation. This author strongly disputes these claims, and instead will show that the inconsistencies claimed by Safonov and Bertram stem solely from a failure to correctly complete a linear transformation of variables, and a concomitant invalid application of well known fluctuation dissipation relations. When used correctly, such fluctuation-dissipation relations most certainly will apply to magnetic (or other physical) systems with dissipation.

## I. INTRODUCTION

In a recent article,[1] Safonov and Bertram (SB) claim "inconsistent" results will occur when applying the "Callen-Welton fluctuation dissipation theorem" to "magnetic systems with dissipation". Since this author's prior work[2,3] appears to be is the real target[4] of this critique, he believes a correction of the record is in order. After brief review of the classical fluctuation-dissipation theorem (FDT), it will be readily shown below (Secs. III, IV) that the aforementioned claims of "inconsistencies" are without merit, and result solely from a failure to complete a proper transformation of variables.

## II. BRIEF REVIEW OF CLASSICAL FTD

For a *linear* system described by a set of real, *macroscopic* dynamical variables $\{x_j(t)\}$, $j=1,2,...N$, along with a set of externally applied forces $\{f_k(t)\}$, $k=1,2,...M$, the response of the system to any individual $f_k(t)$ is expressed as

$$x_j(t) = \int_0^\infty I_{jk}(\tau) f_k(t-\tau) d\tau \Rightarrow \chi_{jk}(\omega) = \int_0^\infty I_{jk}(\tau) e^{i\omega\tau} d\tau \quad (1)$$

The relationship between the impulse responses $I_{jk}(t)$ and the complex susceptibilities $\chi_{jk}(\omega)$ are well known from linear system theory.[5,6] The $\{x_j(t)\}$ are defined such that $\langle x_j \rangle = 0$, where brackets $\langle \ \rangle$ denote a time/ensemble stochastic average *at thermal equilibrium*. There is additionally defined a set of *conjugate* dynamical variables $\{y_{k=1,2,...M}(t)\}$, such that $E = -\sum_{k=1}^M y_k(t) f_k(t)$ is the interaction energy of the forces $f_k$ with the system.

A fundamental, general statement of the classical FDT, proven in the classic paper of Kubo[6], can be expressed in the form[7]:

$$\chi_{jk}(\omega) = \frac{1}{k_B T} \int_0^\infty \langle x_j(\tau) \dot{y}_k(0) \rangle e^{i\omega\tau} d\tau \quad \text{(where } \dot{y}(t) \equiv \frac{dy(t)}{dt}\text{)} \quad (2)$$

$$\Rightarrow \chi_{jk}(\omega) = \langle x_j(0) y_k(0) \rangle + (i\omega/k_B T) \int_0^\infty \langle x_j(\tau) y_k(0) \rangle e^{i\omega\tau} d\tau \quad (3)$$

where $k_B$ is Boltzmann's constant, and $T$ is the equilibrium temperature of the system. Equation (3) follows from the stationary property $d/dt_0 \langle x_j(\tau+t_0) y_k(t_0) \rangle = 0$ of thermal equilibrium.

The author has previously[2] reformulated Eqs. (3) in terms of the measurable components $S_{x_i x_j}(\omega) \equiv \int_{-\infty}^\infty \langle x_j(\tau) x_k(0) \rangle e^{i\omega\tau} d\tau$ of the ($N \times N$) power spectral density matrix[5] describing the stochastic thermal fluctuations of the the set $\{x_{j=1,2..N}\}$. *However, doing so requires restriction to self-conjugate systems with $M=N$, such that*[2,3]

$$E\{x_j\} = -C \sum_{j=1}^N x_j f_j \quad (4)$$

where $C$ is a single, positive scaling constant. *If, and only if* Eq. (4) holds, it was earlier shown that[2]

$$S_{x_i x_j}(\omega) = \frac{k_B T}{C} \frac{\chi_{jk}(\omega) - \chi^*_{kj}(\omega)}{i\omega} \quad (5)$$

The inverse Fourier transform of Eq. 5 yields the following statement of the classical FDT[2,3]:

$$\langle x_j(\tau) x_k(0) \rangle = \frac{k_B T}{C} \int_{-\infty}^\infty \frac{\chi_{jk}(\omega) - \chi^*_{kj}(\omega)}{i\omega} e^{-i\omega\tau} \frac{d\omega}{2\pi} \quad (6)$$

One may attribute the thermal fluctuations of the $\{x_j\}$ as driven by a conjugate set of stationary "random" thermal forces $\{f^{\text{r}}_{j=1,2...N}\}$, distinct from the deterministic applied forces $\{f_j\}$. The stochastic properties of the $\{f^{\text{r}}_j\}$ must be "thermodynamically consistent" with those of $\{x_j\}$. The latter is expressed using Eq. 1 as:

$$\langle x_i(\tau) x_j(0) \rangle = \sum_{k,k'=1}^N \int_0^\infty dt \int_0^\infty dt' I_{ik}(t) I_{jk'}(t') \langle f^{\text{r}}_k(\tau+t'-t) f^{\text{r}}_{k'}(0) \rangle \quad (7)$$

Taking the Fourier transform of Eq. 7, along with the latter half of Eq. 1, then yields[2]

$$S_{f^{\text{r}}_i f^{\text{r}}_j}(\omega) = \sum_{k,k'=1}^N \Lambda_{ik}(\omega) S_{x_k x_{k'}}(\omega) \Lambda^*_{jk'}(\omega) \quad (8)$$

In Eq. 8, the $\Lambda_{ij}(\omega)$ form the $N \times N$ *inverse* susceptibility matrix, which is taken to be nonsingular. Using Eq. 5, in Eq. (8), formally completing the matrix multiplications, and then taking an inverse Fourier transform, yields the final result[3]

$$\langle f_j^r(\tau) f_k^r(0) \rangle = \frac{k_B T}{C} \int_{-\infty}^{\infty} \frac{\Lambda_{kj}^*(\omega) - \Lambda_{jk}(\omega)}{i\omega} e^{-i\omega\tau} \frac{d\omega}{2\pi} \quad (9)$$

Eq. 9 is a second form of the classical FDT, complimentary to the first form described by Eq. 6. However, being derived from Eq. 6, it is not an independent statement. Eqs. (3-9) are identical to results derived previously by this author.[2,3]

### III. DISCUSSION

In the present context, it cannot be over emphasized that Eqs. (6) or (9) *are valid only for systems described in terms of conjugate variables sets* $\{x_j, f_j\}$ *satisfying the self-conjugate constraint of Eq. (4).* The necessity of this restriction, made *explicit* in this authors' previous work,[2,3] can be further understood here using simple but very general algebraic arguments. Consider the compact matrix/vector notation: $\{x_j\} \Leftrightarrow \boldsymbol{x}$, $\sum_j x_j f_j \Leftrightarrow \boldsymbol{f}^T \cdot \boldsymbol{x}$, and $\sum_j A_{ij} x_j \Leftrightarrow A \cdot \boldsymbol{x}$, for $N \times 1$ "column" vector $\boldsymbol{x}$, transposed $1 \times N$ "row" vector $\boldsymbol{f}^T$, and $N \times N$ matrix $A$. Starting from any initial *conjugate* set $\{\boldsymbol{x}, \boldsymbol{f}\}$ satisfying Eqs. (4), one may be transform to an alternative conjugate set $\{\boldsymbol{x}', \boldsymbol{f}'\}$ via the following matrix transformation:

$$\boldsymbol{x}' \equiv U^{-1} \cdot \boldsymbol{x}, \boldsymbol{f}' \equiv U^T \cdot \boldsymbol{f}$$
$$\Rightarrow \boldsymbol{f}'^T \cdot \boldsymbol{x}' = (\boldsymbol{f}^T \cdot U) \cdot (U^{-1} \cdot \boldsymbol{x}) = \boldsymbol{f}^T \cdot \boldsymbol{x} \quad (10)$$
$$\Rightarrow E = -C \boldsymbol{f}^T \cdot \boldsymbol{x} = -C \boldsymbol{f}'^T \cdot \boldsymbol{x}'$$

which manifestly preserves the self-conjugate property of Eq. (4) for the set $\{\boldsymbol{x}', \boldsymbol{f}'\}$, given *any* real, nonsingular, $N \times N$ transformation matrix $U$ with inverse $U^{-1}$ and transpose $U^T$.

Substituting $\boldsymbol{x}(t) = \boldsymbol{x}(\omega) e^{-i\omega t}$ and $\boldsymbol{f}(t) = \boldsymbol{f}(\omega) e^{-i\omega t}$ in Eqs. (1), the frequency domain equations of motion are $\chi(\omega) \cdot \boldsymbol{f}(\omega) = \boldsymbol{x}(\omega)$, or alternatively, $\Lambda(\omega) \cdot \boldsymbol{x}(\omega) = \boldsymbol{f}(\omega)$. Starting from the latter, and using any *conjugate* set $\{\boldsymbol{x}, \boldsymbol{f}\}$, substitution of *both* $\boldsymbol{f}' = U^T \cdot \boldsymbol{f}$ and $\boldsymbol{x} = U \cdot \boldsymbol{x}'$ from Eq. (10) gives $(U^T \cdot \Lambda(\omega) \cdot U) \cdot \boldsymbol{x}'(\omega) = \boldsymbol{f}'(\omega)$. This immediately specifies the proper conjugate transformation for the inverse susceptibility matrix:

$$\Lambda'(\omega) \equiv U^T \cdot \Lambda(\omega) \cdot U. \quad (11a)$$
$$\Rightarrow \Lambda'^\dagger - \Lambda' = U^T \cdot (\Lambda^\dagger - \Lambda) \cdot U \quad (11b)$$

where hermitian transpose $(\Lambda^\dagger)_{jk} \equiv \Lambda_{kj}^*$ Comparing Eqs. (9) and (11b), it follows that the force correlation matrices necessarily transform as

$$\langle \boldsymbol{f}'^r(\tau) \boldsymbol{f}'^{rT}(0) \rangle = U^T \cdot \langle \boldsymbol{f}^r(\tau) \boldsymbol{f}^{rT}(0) \rangle \cdot U \quad (12)$$

where the "outer" product $\boldsymbol{f}^r(\tau) \boldsymbol{f}^{rT}(0)$ is an $N \times N$ matrix

A key observation is that Eq. (12) is identical to that obtained *solely* by direct algebraic substitution with $\boldsymbol{f}'^r = U^T \cdot \boldsymbol{f}^r$. This intrinsic self-consistency between Eqs. (9-11) and Eq. (12) indicates the necessity of the self-conjugate constraint of Eqs. (4), as expressed by the remaining half $\boldsymbol{x}' = U^{-1} \cdot \boldsymbol{x}$ of the conjugate transformation of Eqs. (10).

This is more fully demonstrated as follows. For a given conjugate variable set $\{\boldsymbol{x}, \boldsymbol{f}\}$, one can alternatively form a "one-sided transformation" by left-multiplying equations $\Lambda(\omega) \cdot \boldsymbol{x}(\omega) = \boldsymbol{f}(\omega)$ by an *arbitrary* nonsingular $U^T$. This generates an infinity of forces $\boldsymbol{f}' = U^T \cdot \boldsymbol{f}$ which *do satisfy* Eq. (12), and a corresponding infinite variety of valid equations of motion corresponding to

$$\tilde{\Lambda}(\omega) \cdot \boldsymbol{x}(\omega) = \boldsymbol{f}'(\omega), \text{ where. } \tilde{\Lambda}(\omega) \equiv U^T \cdot \Lambda(\omega) \quad (13a)$$
$$\Rightarrow \boldsymbol{f}'^T \cdot \boldsymbol{x} = \boldsymbol{f}^T \cdot U \cdot \boldsymbol{x} \neq \boldsymbol{f}^T \cdot \boldsymbol{x} \quad (13b)$$
$$\Rightarrow \tilde{\Lambda}^\dagger - \tilde{\Lambda} = \Lambda^\dagger \cdot U - U^T \cdot \Lambda \neq U^T \cdot (\Lambda^\dagger - \Lambda) \cdot U \quad (13c)$$

However, despite the validity of Eqs. (13a), Eqs. (13b) explicitly show that the *non-conjugate set* $\{\boldsymbol{x}, \boldsymbol{f}'\}$ will *not* satisfy Eqs. (4), invalidating use of the corresponding $\tilde{\Lambda}(\omega)$ in Eqs. (9). Indeed, Eqs. (13c) explicitly show that improper use of $\tilde{\Lambda}(\omega)$ in Eqs. (9) is *guaranteed* in general to produce erroneous results that are *necessarily inconsistent* with those of Eqs. (12). This form of argument applies equally to the inverted equations $\chi(\omega) \cdot \boldsymbol{f}(\omega) = \boldsymbol{x}(\omega)$, with analogous restricted use of conjugate variables to obtain meaningful results when using Eqs. (6).

### IV. FDT AND MAGNETIC DAMPING

Claims of "inconsistency" by SB[1] in regards to Eqs. (9) result *solely* from an invalid use of an aforementioned "one-sided transformation" which fails to properly complete the transformation to a *conjugate* variable set. The specific example used in Sec. III of SB[1] is that of the small-signal motion of transverse magnetization components $\boldsymbol{m}(t) \equiv (m_x(t), m_y(t))$ for a single domain magnetic particle whose equilibrium magnetization lies along the $\hat{z}$-axis, and which is subject to a "small" transverse perturbation field $\boldsymbol{h}(t) \equiv (h_x(t), h_y(t))$. From the free energy expression

$$E_{\text{free}} / M_s V = (\tfrac{1}{2} H_x m_x^2 + \tfrac{1}{2} H_y m_y^2) - \boldsymbol{h}(t) \cdot \boldsymbol{m}(t) \quad (14)$$

the rightmost Zeeman interaction term in Eq. 14 is identical in form to that of Eq. (4), with $\{x_j\} \leftrightarrow \boldsymbol{m}$, $\{f_j\} \leftrightarrow \boldsymbol{h}$, and $C \leftrightarrow M_s V$. *It follows that* $\{\boldsymbol{m}, \boldsymbol{h}\}$ *form a proper conjugate set.*

The arguments in SB[1] center around a comparison of two alternative equations of motion, namely those of the Landau-Lifshitz-Gilbert[8] (LLG) or Bloch-Bloembergen[9] (BB) form

$$\text{LLG:} \quad \frac{1}{\gamma} \begin{pmatrix} \alpha & -1 \\ 1 & \alpha \end{pmatrix} \cdot \begin{pmatrix} \dot{m}_x \\ \dot{m}_y \end{pmatrix} + \begin{pmatrix} H_x & 0 \\ 0 & H_y \end{pmatrix} \cdot \begin{pmatrix} m_x \\ m_y \end{pmatrix} = \begin{pmatrix} h_x \\ h_y \end{pmatrix} \quad (15)$$

$$\text{BB:} \quad \frac{1}{\gamma} \begin{pmatrix} 0 & -1 \\ 1 & 0 \end{pmatrix} \cdot \begin{pmatrix} \dot{m}_x \\ \dot{m}_y \end{pmatrix} + \begin{pmatrix} H_x & -1/\gamma\tau_2 \\ 1/\gamma\tau_2 & H_y \end{pmatrix} \cdot \begin{pmatrix} m_x \\ m_y \end{pmatrix} = \begin{pmatrix} h_x \\ h_y \end{pmatrix} \quad (16)$$

With $\boldsymbol{m}(t) \to \boldsymbol{m}(\omega)e^{-i\omega t}, \boldsymbol{h}(t) \to \Lambda(\omega) \cdot \boldsymbol{m}(\omega)e^{-i\omega t}$, the $2\times 2$ inverse susceptibility matrices $\Lambda(\omega)$ are readily found from inspection of Eqs. (15,16) to be

$$\Lambda_{\text{LLG}}(\omega) = \frac{1}{\gamma}\begin{pmatrix} \gamma H_x - i\alpha\omega & i\omega \\ -i\omega & \gamma H_y - i\alpha\omega \end{pmatrix} \quad (17)$$

$$\Lambda_{\text{BB}}(\omega) = \frac{1}{\gamma}\begin{pmatrix} \gamma H_x & i\omega - 1/\tau_2 \\ 1/\tau_2 - i\omega & \gamma H_y \end{pmatrix} \quad (18)$$

Application of the FDT of Eq. (9) to $\Lambda_{\text{LLG}}(\omega)$ in Eq. (17) readily yields

$$\begin{pmatrix} \langle h_x^r(t)h_x^r(0)\rangle & \langle h_x^r(t)h_y^r(0)\rangle \\ \langle h_y^r(t)h_x^r(0)\rangle & \langle h_y^r(t)h_y^r(0)\rangle \end{pmatrix}_{\text{LLG}} = \frac{2\alpha k_B T}{\gamma M_s V}\begin{pmatrix} 1 & 0 \\ 0 & 1 \end{pmatrix}\delta(t) \quad (19)$$

which agrees identically with classic result of Brown.[10] The left side of Eq. (19) will henceforth be denoted using the simplified notation, $\langle \boldsymbol{h}^r(t)\boldsymbol{h}^r(0)\rangle$.

In stark contrast to $\Lambda_{\text{LLG}}(\omega)$, it is seen from Eq. (18) that $\Lambda_{\text{BB}}^T(\omega \to 0) \neq \Lambda_{\text{BB}}(\omega \to 0)$ is *asymmetric*. This nonphysical property of BB damping (also noted previously[3,11]) can be shown[12] to violate conservation of energy. Ignoring this, application of Eq. (9) to Eqs. (18) yields the manifestly nonphysical result

$$\langle \boldsymbol{h}^r(t)\boldsymbol{h}^r(0)\rangle_{\text{BB}} = \frac{2k_B T/\tau_2}{\gamma M_s V}\begin{pmatrix} 0 & -1 \\ 1 & 0 \end{pmatrix}\text{sgn}(t) \quad (20)$$

SB[1] introduce a simple change of field variables $\boldsymbol{h}(t) \to \boldsymbol{h}'(t)$, which (ignoring terms of order $\alpha^2$, as do SB[1]), is defined in the notation of Eq. (10) as

$$\begin{pmatrix} h'_x \\ h'_y \end{pmatrix} \equiv \begin{pmatrix} 1 & -\alpha \\ \alpha & 1 \end{pmatrix}\cdot\begin{pmatrix} h_x \\ h_y \end{pmatrix} \Leftrightarrow U \equiv \begin{pmatrix} 1 & \alpha \\ -\alpha & 1 \end{pmatrix} \quad (21)$$

It follows by direct substitution, and/or from Eqs. (12) that

$$\langle \boldsymbol{h}'^r(t)\boldsymbol{h}'^r(0)\rangle_{\text{LLG}} = \frac{2\alpha k_B T}{\gamma M_s V}\begin{pmatrix} 1 & 0 \\ 0 & 1 \end{pmatrix}\delta(t). \quad (22)$$

which (to order $\alpha$) is the *same* as $\langle \boldsymbol{h}^r(t)\boldsymbol{h}^r(0)\rangle_{\text{LLG}}$. In the isotropic case $H_x = H_y \to H_0$, SB[1] re-express Eqs.(15) in terms of $\boldsymbol{m}(t)$ and $\boldsymbol{h}'(t)$, the strict equivalent of left-multiplying both sides of Eqs. (15) by matrix $U^T$ of Eq. (21). To first order in $\alpha$:

$$\text{LLG: } \frac{1}{\gamma}\begin{pmatrix} 0 & -1 \\ 1 & 0 \end{pmatrix}\cdot\begin{pmatrix} \dot{m}_x \\ \dot{m}_y \end{pmatrix} + \begin{pmatrix} H_0 & -\alpha H_0 \\ \alpha H_0 & H_0 \end{pmatrix}\cdot\begin{pmatrix} m_x \\ m_y \end{pmatrix} = \begin{pmatrix} h'_x \\ h'_y \end{pmatrix} \quad (23)$$

In reference to the discussion of one-sided, *non-conjugate* transformations described by Eqs. (13), one finds from inspection of Eqs. (23) that

$$\tilde{\Lambda}_{\text{LLG}}(\omega) = \frac{1}{\gamma}\begin{pmatrix} \gamma H_0 & i\omega - \alpha\gamma H_0 \\ \alpha\gamma H_0 - i\omega & \gamma H_0 \end{pmatrix} \quad (24)$$

SB[1] go on to finally argue that the *difference* in the results for $\langle \boldsymbol{h}'^r(t)\boldsymbol{h}'^r(0)\rangle_{\text{LLG}}$ in Eqs. (22) and that for $\langle \boldsymbol{h}^r(t)\boldsymbol{h}^r(0)\rangle_{\text{BB}}$ in Eqs (20) must constitute an "inconsistency" of the FDT statement of Eq. (9), because BB Eqs. (16,18) expressed in $\{\boldsymbol{m},\boldsymbol{h}\}$ are identical in form to LLG Eqs. (23,24) expressed in $\{\boldsymbol{m},\boldsymbol{h}'\}$.

The fatal flaw of this argument was described at the end of Sec. III. Namely, the $\{\boldsymbol{m},\boldsymbol{h}'\}$ in Eqs. (23) constitute a *non-conjugate* variable set, for which the application of Eq. (9) using the resultant $\tilde{\Lambda}_{\text{LLG}}(\omega)$ of Eqs. (24) was and is *manifestly invalid*. In particular, it follows from Eqs. (14) and (21) that

$$E_{\text{Zeeman}}/M_s V = -\boldsymbol{m}\cdot\boldsymbol{h} = -(\boldsymbol{m}\cdot\boldsymbol{h}' + \alpha(\boldsymbol{m}\times\boldsymbol{h}')\cdot\hat{\boldsymbol{z}}) \quad (25)$$

Comparing with Eq. (4), $\{\boldsymbol{m},\boldsymbol{h}'\}$ are not only *non*-conjugate, but the error term in Eq. (25) is of *first* order in $\alpha$.

If one instead simply completes the *conjugate* transformation $\{\boldsymbol{m},\boldsymbol{h}\} \to \{\boldsymbol{m}',\boldsymbol{h}'\}$ via Eqs. (10), and replaces $\boldsymbol{m} \to U\cdot\boldsymbol{m}'$ in Eqs. 23, one easily finds (ignoring terms of order $\alpha^2$) that

$$\text{LLG: } \Rightarrow \frac{1}{\gamma}\begin{pmatrix} \alpha & -1 \\ 1 & \alpha \end{pmatrix}\cdot\begin{pmatrix} \dot{m}'_x \\ \dot{m}'_y \end{pmatrix} + \begin{pmatrix} H_0 & 0 \\ 0 & H_0 \end{pmatrix}\cdot\begin{pmatrix} m'_x \\ m'_y \end{pmatrix} = \begin{pmatrix} h'_x \\ h'_y \end{pmatrix} \quad (26)$$

which is identical in form with Eqs.(15). Application of the FDT of Eq. (9) to Eqs. (26) will thus yield *fully consistent* results when compared with Eqs. (22).

It more generally follows from Eqs. (11b) that the proper conjugate transformation $\Lambda' \equiv U^T\cdot\Lambda\cdot U \Leftrightarrow \Lambda'^T \equiv U^T\cdot\Lambda^T\cdot U$ manifestly preserves the symmetry (i.e., $\Lambda'^T = \Lambda'$ if $\Lambda^T = \Lambda$) or *asymmetry* of $\Lambda(\omega)$. Hence, $\Lambda'_{\text{LLG}}(\omega \to 0)$ will always be symmetric, and $\Lambda'_{\text{BB}}(\omega \to 0)$ will remain unphysically asymmetric regardless of the choice of transformation matrix $U$. Indeed, the premise in SB[1] that the difference in the basic physics of the LLG and BB damping terms could be eliminated by means of a simple variable transformation was and is transparently erroneous. In summary, the claims of "inconsistency" leveled by SB[1] against this author's formulation[4] of the FDT have no basis in reality, and are solely and demonstrably a product of their failure to complete a simple variable transformation to a proper set of self-conjugate variables prior to application of the general FDT relations of Eqs. (9).

## V. SOME FURTHER REMARKS

The self-conjugate constraint of Eq. 4 is equivalently stated in the very first equation in Sec. II of SB[1]. However, for reasons unknown, this constraint is simply "forgotten" in Sec III of SB[1] where the aforementioned erroneous arguments concerning "inconsistencies" are given.

The derivation of Eqs. 3-9 are those of this author. Results virtually *identical* to Eq. 6 and Eq.9 (with $C = 1$) appear in SB[1], which SB repeatedly claim to originate from the work of "Callen-Welton".[13] This claim is inaccurate,[4,14] since both the initial[2] and subsequent[3] results of this author were *explicitly* demonstrated[2] as being derived from the the Kubo statement[6,7] of the classical FDT (Eq. 2). SB's repeated criticism[14] of the Callen-Welton[13] formulation of of the FDT is hence irrelevant to the discussion of alleged "inconsistencies" in Eqs. (9). Nonetheless, this author finds SB's criticism of "Callen-Welton" to also be unjustified.[15]

Contrary to claims in SB[1] (which may apply elsewhere[14]), the derivation steps of Eqs. (7-8) in going from Eq. (6) to Eq. (9) do *not* assume any direct relationship between stochastic variables $x(t)$ and $f^r(t)$ or their Fourier transforms, but deal *only* with proper stochastic-averaged time correlations and/or power spectral densities of these quantities.

SB[1] claim that Eq. (9) implies a "physically impossible" dependence of the noise correlations of the $f^r(t)$ with physically unrelated properties of the dynamic system described by $\Lambda(\omega)$. However, Eq. (9) depends *only* on the anti-hermitian, or *dissipative* component of $\Lambda(\omega)$. For example, the dynamics described by Eqs. (15,17) depends primarily on stiffness fields $H_{x,y}$, but the noise correlations described in Eq. (19) depend *only* on the damping parameter $\alpha$.

SB characterize the results of Eqs. 9 as an "approximation", strictly valid only in the case of systems "without dissipation".[15] In contrast to the arguments of Landau-Lifshitz,[13,14] Kubo's proof of the FDT formally expresses all *macroscopic* variables $x_j$ in terms of a *complete* set $\{p,q\}$ of *microscopic/atomistic* coordinates. It is the dynamics of the $\{p,q\}$ which are then treated via Hamiltonian mechanics (classical or quantal). At the "atomic" level, macroscopic dissipation has no meaning, there being only energy exchanging interactions among the internal $\{p,q\}$ degrees of freedom of the system, or between the system and its environment (thermal bath). Interactions between the system and environment are included by Kubo via the statistical mechanics of the canonical ensemble, which conserves temperature, but *not energy* of the system. The flow of energy from the system to its environment is the physical origin of all dissipation processes. However, the final Kubo statement of the FDT in Eq. (2) makes reference *only* to macroscopic variables $x_j$, and their conjugate forces (implicitly through the $y_k$ and the $\chi_{jk}$). The relationship between the latter is described by *macroscopic* equations of motion (time or frequency domain) which will necessarily contain terms describing macroscopic dissipation (or "damping") properties. The Gilbert equations of motion given by Eqs. (15,17), which contain the damping parameter $\alpha$, are but one such example. The critique of SB[1] that fluctuation-*dissipation* relations such as Eqs. (6) or (9) will *not* apply to systems "with dissipation" is manifestly incorrect.[7,15]

## REFERENCES

[1]V. L Safonov and H. N. Bertram, Phys. Rev. B **71**, 224402 (2005).

[2]N. Smith, J. Appl. Phys. **90**, 5768 (2001).

[3]N. Smith, J. Appl. Phys. **92**, 3877 (2002).

[4]A "reply" to Ref. 3, posted in 2002 by Safonov and Bertram, arXiv:cond-mat/0211147, raises similar "inconsistency" arguments as appear in SB[1], *but makes no reference whatsoever* to the work of Callen-Welton[13].

[5] A. Papoulis, *Probability, Random variables, and Stochastic Processes*, 2$^{nd}$ ed. (McGraw-Hill, New York, 1984). Chaps. 9 and 10.

[6] R. Kubo, Rep. Prog. Phys. **29**, 255 (1966).

[7]Equations. (2) of this article is the same as Eq. (5.21) of Kubo[6]. The reader is welcomed to consider the extensive role of friction and dissipation in the FDT as discussed by Kubo for a system consisting of a Brownian motion particle.

[8]T. L. Gilbert, Phys. Rev. **100**, 1243 (1955).

[9]. Bloembergen, Phys. Rev. **78**, 572 (1950).

[10]W. F. Brown, Phys. Rev. **130**, 1677 (1963). (Eq. (3.26) with $\eta \equiv \alpha/\gamma M_s$.)

[11]V. Cambersky and C. E. Patton, Phys. Rev. B, **11**, 2668 (1975).

[12]For linear systems described with a *self-conjugate* variable set $\{x_j, f_j\}$, it follows from Eqs. (4) that $\Delta W = C\sum_j \int f_j dx_j$ is the work done by forces $\{f_j\}$ on the system. For isothermal, *quasi*-static changes at sufficiently low (but nonzero) frequency such that $\dot{x}_k \to 0$ and $f_j \to \sum_k \Lambda_{jk}(\omega \to 0) x_k$, then $\Delta W \to E(\{x\}) = \frac{C}{2}\sum_{j,k} x_j \Lambda_{jk}(\omega \to 0) x_k$ is the free energy, a thermodynamic *state-function* with minimum at $\{x_j\} = 0$ (since $\langle x_j \rangle = 0$). More generally, $\Lambda_{jk}(\omega \to 0) = 1/C \, \partial^2 E/\partial x_j \partial x_k = \Lambda_{kj}(\omega \to 0)$ must be *symmetric* positive definite, or else violate the first law of thermodynamics. See also Ref. 8, Sec. 125; and L. D. Landau, E. M. Lifshitz, and L. P. Pitaevskii, *Electrodynamics of Continuous Media*, 2$^{nd}$ ed. (Pergamon Press, New York 1984). Chap. 4, Sec. 29.

[13]H. B. Callen and T. A. Welton, Phys. Rev. **83**, 34 (1951). Though repeatedly referred to by SB[1] as "Callen-Welton", the actual description of the FDT in Sec. II of SB[1] much more closely follows that given by L. D. Landau and E. M. Lifshitz, *Statistical Physics* (Pergamon, New York, 1980), Part I, Chap. 12

[14] The more rigourous derivation of the FDT given by Kubo[6] differs from that of Callen-Welton/Landau-Lifshitz,[13] which is based on first order quantum mechanical time dependent perturbation theory. Possible shortcomings in the latter cited by SB[1] have no bearing on the validity of Eqs. (2-9) in this article, regardless that some final results of Landau-Lifshitz[13] are similar in form to the Fourier transforms of Eqs. (6,9).

[15]The claims in SB[1] concerning the non-applicability of the work of Callen-Welton[13] to systems with dissipation appears to be motivated by similar claims of Y. L. Klimontovich, Nonlinear Phenomena in Complex Systems **5**, 372 (2002). The Klimontovich argument objects to the additional frequency dependence from the factor $k_B T \to (\hbar\omega/2)\coth(\hbar\omega/2k_B T)$ that appears in expressions[13] for the power spectral density $S(\omega)$ *in the quantum limit* $\hbar\omega/k_B T \geq 1$. Since discussions here and in SB[1] are restricted to the *classical* FDT, the citation of Klimontovich by SB[1] in this context is at best irrelevant, if not misleading. In this author's view, the Klimontovich arguments are also suspect due to a lack of consideration of the quantum non-commutability, e.g., $\langle x(\tau)x(0)\rangle \neq \langle x(0)x(\tau)\rangle$ of the correlations used to define $S(\omega)$. Classical$\to$quantum correspondence rules are considered in detail by Kubo[6]. In particular, Kubo's "canonical" correspondence $\langle x(\tau)x(0)\rangle \to \langle 1/\beta \int_0^\beta e^{\lambda H} x(\tau) e^{-\lambda H} x(0) d\lambda \rangle$, where $\beta \equiv 1/k_B T$ and $H\{p,q\} \equiv$ Hamiltonian, removes the $k_B T \to (\hbar\omega/2)\coth(\hbar\omega/2k_B T)$ substitution, leaves FDT expressions otherwise unchanged from their classical form, and obviates the arguments of Klimontovich.